\documentstyle[12pt]{article}

\newcommand{\mabel}[1]{\label{#1}}
\newcommand{\mibitem}[1]{\bibitem{#1}}

\newcommand{\Kac}{Ka\v c\ }
\newcommand{\halfxxz}{spin $\half$ XXZ chain}
\newcommand{\eq}{Equation }
\newcommand{\BAE}{Bethe Ansatz equations}
\newcommand{\QISM}{Quantum Inverse Scattering Method}
\newcommand{\ABA}{algebraic Bethe Ansatz}
\newcommand{\BA}{Bethe Ansatz}
\newcommand{\FSC}{finite size corrections}
\newcommand{\CFT}{conformal field theory}
\newcommand{\CFTs}{conformal field theories}
\newcommand{\FCR}{fundamental commutation relations}

\newcommand{\be}{\begin{equation}} 
\newcommand{\ee}{\end{equation}} 
\newcommand{\bea}{\begin{eqnarray}} 
\newcommand{\eea}{\end{eqnarray}} 
\newcommand{\beastar}{\begin{eqnarray*}} 
\newcommand{\eeastar}{\end{eqnarray*}} 

\renewcommand{\.}{\ .}
\renewcommand{\,}{\ ,}
\renewcommand{\;}{\ ;\ \ \ \ }
\newcommand{\refe}[1]{(\ref{#1})}
\newcommand{\vali}{\vspace{4mm}} 
\newcommand{\mm}{\hspace{0.5mm}} 
\newcommand{\mmplus}{\mm + \mm}

\newcommand{\NP}{Nucl. Phys. }
\newcommand{\PRL}{Phys. Rev. Lett. }

\newcommand{\PL}{Phys. Lett. }

\newcommand{\CMP}{Commun. Math. Phys. }

\newcommand{\MPL}{Mod. Phys. Lett. }

\newcommand{\sss}[1]{{{\scriptscriptstyle #1}}} 
\newcommand{\kpert}{{\textstyle \frac{K}{2}}} 
\newcommand{\Kmiopert}{{{\scriptscriptstyle \frac{1}{2}(K-1)}}}
\newcommand{\Lmiopert}{{{\scriptscriptstyle \frac{1}{2}(L-1)}}}
\newcommand{\jhalf}{{ \frac{1}{2}}\mm} 

\newcommand{\half}{{\textstyle \frac{1}{2}}\mm}
\newcommand{\quarter}{{\textstyle \frac{1}{4}}\mm}
\newcommand{\ipert}{{\textstyle \frac{{\rm i}}{2}}\mm}
\newcommand{\operN}{{\textstyle \frac{1}{N}}} 

\newcommand{\pipert}{{\textstyle \frac{\pi}{2}}} 
\newcommand{\piperf}{{\textstyle \frac{\pi}{4}}} 

\newcommand{\opergam}{{\textstyle \frac{1}{\gamma}}\mm}
\newcommand{\opertgam}{{\textstyle \frac{1}{2\gamma}}} 
\newcommand{\gampert}{{\textstyle \frac{\gamma}{2}}\mm}
\newcommand{\pipergam}{{\textstyle \frac{\pi}{\gamma}}} 
\newcommand{\gamperpi}{{\textstyle \frac{\gamma}{\pi}}} 

\newcommand{\singam}{\sin\gamma}

\newcommand{\til}{\tilde}
\newcommand{\Bardelta}{\bar\Delta} 
\newcommand{\sK}{{\sss K}} 
\newcommand{\sL}{{\sss L}} 
\newcommand{\sN}{{\sss N}} 

\newcommand{\kap}{\kappa} 
\newcommand{\Lam}{\Lambda}

\renewcommand{\i}{\mm{\rm i}\mm} 

\newcommand{\A}{{\cal A}} 
\newcommand{\B}{{\cal B}} 
\newcommand{\C}{{\cal C}} 
\newcommand{\D}{{\cal D}} 
\renewcommand{\H}{{\cal H}} 
\renewcommand{\L}{{\cal L}} 
\renewcommand{\O}{{\cal O}} 
\renewcommand{\P}{{\cal P}} 
\renewcommand{\S}{{\cal S}} 

\newcommand{\gam}{\gamma}
\newcommand{\lam}{\lambda}
\newcommand{\tilgam}{{\tilde\gamma}}
\newcommand{\tillam}{{\tilde\lambda}}

\newcommand{\longat}[1]{\hskip 0.5mm {\rule[-4mm]{.15mm}{8mm}}_{\hskip
0.5mm #1}} 

\newcommand{\tensor}{\otimes}
\newcommand{\del}{\partial}

\newcommand{\dalembert}{\mm\Box\mm}

\newcommand{\e}[1]{\ {\rm e}^{ #1 }\mm}

\newcommand{\mat}[2]{\left(\begin{array}{#1} #2 \end{array}\right)}

\newcommand{\BAEspspl}[2]{\frac{\sinh\left(#1\mm+\mm #2\right)
	}{\sinh\left(#1\mm -\mm #2\right)}\mm} 
\newcommand{\BAEspsmi}[2]{\frac{\sinh\left(#1\mm-\mm #2\right)
	}{\sinh\left(#1\mm +\mm #2\right)}\mm} 

\newcommand{\BAEsps}[2]{\frac{\sinh (#1 + #2)}{\sinh (#1 - #2)}} 

\newcommand{\mmdots}{\mm\ldots\mm}
\renewcommand{\iff}{\ \Longleftrightarrow\ }
\newcommand{\all}{\ \ \ \ \forall\ }
\newcommand{\slqt}{$U_q(sl_2)$}
\renewcommand{\Im}{\mm \mbox{Im}\mm}
\renewcommand{\Re}{\mm \mbox{Re}\mm}

\newcommand{\com}[2]{\mm \left\lbrack #1 ,\mm\mm #2 \right\rbrack\mm}
\newcommand{\pois}[2]{\mm\left\{ #1 ,\mm\mm  #2 \right\}\mm} 
\newcommand{\set}[1]{\{#1\}} 

\newcommand{\mod}{\ \mbox{mod}\ }

\newcommand{\trac}[2]{{\textstyle\frac{#1}{#2}}}
\newcommand{\stack}[2]{\stackrel{{\sss #1}}{#2}} 
\newcommand{\unit}{1\!\!\hskip 1pt\mbox{l}}

\newcommand{\sigmaI}{\mat{rr}{0&1\\1&0}}

\begin{document}

\baselineskip 17pt plus 2pt minus 1pt




\begin{titlepage}

\begin{flushright}
{ HU-TFT-95-15\\
hep-th/9506023
}
\end{flushright}

\vskip 3mm

\begin{center}
{\large \bf Connections of the Liouville Model and XXZ Spin Chain\\}
\end{center}

\vskip 9mm

\begin{center}
{\bf Ludvig D. Faddeev$^*$}\\
\vskip 1mm
{\it St Petersburg Branch of Steklov Mathematical Institute,\\
Fontanka 27, St Petersburg 191011, Russia} \\
and\\
{\it Research Institute for Theoretical Physics, University of
Helsinki \\
Siltavuorenpenger 20 C, SF-00170 Helsinki, Finland  }

\vskip 9mm

{\bf Olav Tirkkonen$^{**}$  } \\
\vskip 1mm
{\it Research Institute for Theoretical Physics, University of
Helsinki \\
Siltavuorenpenger 20 C, SF-00170 Helsinki, Finland } \\
and \\
{\it Department of Physics, University of British Columbia \\
6224 Agricultural Road, Vancouver, BC, V6T 1Z1 Canada}

\vskip 13mm

\rm

{\bf Abstract}

\vskip 1mm

\end{center}
\noindent
The quantum theory of the Liouville model with imaginary field is
considered using the \QISM. An integrable structure with nontrivial
spectral parameter dependence is developed for lattice Liouville
theory by scaling the $L$-matrix of lattice sine-Gordon theory. This
$L$-matrix yields \BAE\ for Liouville theory, by the methods of the
\ABA.  Using the string picture of exited Bethe states, the lattice
Liouville Bethe equations are mapped to the corresponding \halfxxz\
equations.  The well developed theory of \FSC\ in spin chains is used
to deduce the conformal properties of the lattice Liouville Bethe
states.  The unitary series of \CFTs\ emerge for Liouville couplings
of the form $\gam = \pi\trac{\nu}{\nu+1}$, corresponding to root of
unity XXZ anisotropies.  The Bethe states give the full spectrum of
the corresponding unitary \CFT, with the primary states in the \Kac
table parameterized by a string length $K$, and the remnant of the
chain length mod $(\nu+1)$.

\vfill

\begin{flushleft}
\rule{5.1 in}{.007 in}\\
$^{*}$ {\small E-mail: faddeev@lomi.spb.su \\ }
$^{**}$ {\small E-mail: tirk@physics.ubc.ca \\ }
\end{flushleft}

\end{titlepage}

\section{Introduction}

Ever since the seminal observation by Polyakov \cite{polyakov} that
Liouville field theory describes two-dimensional gravity as coupled to
a relativistic string, much attention has been devoted to the problem
of quantizing it, see
e.g. \cite{gernev,curtho,fadtak,seiberg,gervais,takhtajan} and
references therein.

Despite the fact that all the arsenals of canonical quantization
\cite{gernev,curtho} and path integral quantization
\cite{seiberg,takhtajan} have been used, much remains unclear about
the quantum theory, especially in the strongly coupled regime with
central charge $1< c < 25$. Some encouraging results have been
acquired for this region utilizing the underlying \slqt\ symmetry of
the quantum theory \cite{gervais}, and by solving the conformal Ward
identities \cite{takhtajan}.

Liouville theory being a completely integrable model, the powerful
methods of the \QISM\ \cite{fadlhou82} should be applicable to
quantizing it. In the spirit of this method, Liouville theory was put
on lattice in \cite{fadtak}, but due to peculiarities of the
integrable structure of Liouville theory, only partial quantum results
were achieved. The concept of lattice Liouville theory, and the
parallel concept of lattice Virasoro algebra were further developed
in \cite{volkov,fadvol,babelon}.

In this paper we take a somewhat different approach to putting
Liouville on the lattice, more suitable for applying the methods of
the \ABA\ (see \cite{fadBA} for recent reviews).  Our
conventions for classical continuum Liouville theory are explained in
{\bf section 2}.

In order to use the full power of the \QISM, one needs a $L$-operator
which depends non-trivially of the spectral parameter. The
$L$-operators used in quantum Liouville theory have so far lacked this
property. In {\bf Section 3} we shall find a remedy for this
shortcoming, and thus the methods of the \ABA\ are in our use.

Following the line of thought of \cite{izekor}, we find in {\bf Section
4} a pseudovacuum for the product of $L$-operators on two adjacent
sites, and derive \BAE\ for lattice Liouville theory. The equations
can be regarded  as  \BAE\ for a XXZ spin chain with spin
$(-\half)$, with an extra phase factor depending on the length of the
chain $N$. More exactly, the extra phase is related to the $N$:th
power of the XXZ anisotropy $q=\e{\i\gam}$, where $\gam$ is the
Liouville coupling constant.

In {\bf Section 5} we then map the spin $(-\half)$ Bethe equations to
the paradigmatic spin $(+\half)$ ones with an extra phase factor. This
is done in the string approach \cite{taksuz} to exited states in the
thermodynamic \BA.  The mapping from Liouville to XXZ is successful
only for certain root of unity anisotropies $q$; the Liouville
coupling has to be of the form $\gam=\pi\trac{\nu}{\nu+1}$, with $\nu$
an integer.  There is a reciprocal one to one correspondence between
Bethe states in the lattice Liouville model and the \halfxxz; highest
strings are mapped to 1-strings and vice versa.

In \cite{blocani,affleck,cardy} it was shown that conformal properties
of two-dimensional statistical models at criticality can be extracted
{}from the finite (but large) size corrections to the eigenvalues of the
transfer matrix. Based on this, systematic methods for analytical
calculation of the \FSC\ have been developed
\cite{devewoy,karowski,kluwezi,boizre}.  For a spin chain or lattice
model, the finite size analysis yields $\operN$ corrections to the
eigenvalues of the transfer matrix, where $N$ is the number of lattice
sites.

The calculation by Karowski \cite{karowski} of the conformal weights
of string exited states in six-vertex and related Potts models will be
of particular interest to us. In {\bf Section 6} we review these
results to for the $\operN$ corrections to a \halfxxz\ with extra
phase factor.

Finally, in {\bf Section 7} we interpret the results in terms of
lattice Liouville theory. As the extra phase factor,
$\exp\{N\pi\trac{\nu}{\nu+1}\}$ is a $\nu+1$:th root of unity, the
thermodynamic limit $N\to\infty$ of our system of \BAE\ should be
taken in steps of $\nu+1$. The phase factor thus becomes a function of
the remnant $\kap = N/2 \mod (\nu+1)$.

The finite size results show that different remnants correspond to
different exited states, thus generalizing the property of
spin chains that chains of even and odd length have different spectra.
The scaling properties of the antiferromagnetic XXZ vacuum with $\kap=1$
yield the central charges of the minimal models \cite{bpz}
belonging to the unitary series of Friedan Qiu and Shenker \cite{fqs}.

For $\kap=0$ we get an ``unrestricted'' sector with central charge $c=1$.
Excluding this sector from the theory corresponds exactly to the RSOS
reduction of a critical SOS model \cite{anbafo} . The spin $\half$
Bethe equations with the extra phase  are exactly the Bethe
equations of a SOS model at criticality. The restriction of these
models is known to produce unitary \CFTs\ with $c<1$
\cite{huse}.

It is very plausible that the exclusion of the $\kap=0$ sector of the
Hilbert space of the theory should be connected to an unitarity
analysis in terms of ``good'' representations of the underlying
\slqt\ symmetry of the theory, c.f. \cite{jutkar}.

With the result of \cite{karowski} that XXZ primary states emerge from
a single string exited over the antiferromagnetic vacuum, we then have
two integers to parameterize exited states: the string length $K$ and
the remnant $\kap$.  Working in the restricted sector, we recover from
the \FSC\ the conformal weights corresponding to the whole \Kac table
of primary states in unitary \CFTs, parameterized by these two
integers.

We view the results announced here as encouraging when it comes to the
analysis of quantum Liouville theory. Generalizing our approach to a
wider class of Liouville coupling constants, possibly along the lines of
\cite{gervais}, might shed more light on the evasive strong coupling regime.

In addition, doing  the   mapping  of Section 5 in the inverse
direction, we see how the Bethe equations of a host of critical
statistical models (six-vertex, Potts, III/IV-critical RSOS) can be
mapped to the lattice Liouville ones. We regard this as an explicit
proof of the conformal invariance of these theories at criticality,
which gives a possibility to find a Lagrangian description of the
corresponding critical field theories.

\section{Classical Liouville theory}

We shall be quantizing Liouville theory on a Minkowskian cylinder,
with the basic field $\Phi(x,t)$.  The space coordinate is periodic,
$x\in[0, 2\pi]$, and time is non-compact, as usual: $t \in [-\infty,
\infty]$.  We take the classical Liouville action in the form
\be
 \S= \opertgam \int_0^{2\pi} dx\mm\left\{\half \dot\Phi^2 - \half \Phi'^2
 - 2 \e{-2 \i \Phi}   + 2\i \Phi''\right\} \.
 \mabel{claction}
\ee
As usual,  $\dot\Phi \equiv \frac{\del}{\del t} \Phi(x,t)$ and $\Phi' \equiv
\frac{\del}{\del x} \Phi(x,t)$.  The so called conformal improvement
term $\i \Phi''$ is required to make the conformal invariance
manifest. It can be considered as the flat space residue of the
coupling to the scalar curvature $\sqrt{g} \mm R(g)\mm\Phi$ of the
Liouville action on a generic Riemann surface, see
e.g. {\cite{seiberg}}.

As it stands, the action of \eq{\refe{claction}} seems very
non-unitary, but as we shall see in the course of this paper, and as
can be inferred from {\cite{gernev,curtho}}, for specific values of
the coupling $\gamma$ the complexities conspire  to yield an
unitary theory after quantization.

The equation of motion corresponding to {\refe{claction}} is
Liouville's equation (with an imaginary field),
\be
 \ddot\Phi - \Phi'' - 4 \i \e{-2\i\Phi} = 0 \.
 \mabel{lioueq}
\ee
 In our parameterization the coupling constant does not appear in the
equation of motion, it only multiplies the action.  By redefining the
field it is easy to recover the usual way
\cite{gernev,curtho,seiberg} of having the coupling in the
exponential.

To unravel the conformal invariance of the theory, it is easiest to move
over to the Hamiltonian picture.  Taking the conjugate momentum to be
$\Pi = \half \dot\Phi$, with Poisson Brackets
\be
 \{ \Phi(x),\mm \Pi(y) \} = - \gamma \mm\delta(x-y) \,
 \mabel{pbs}
\ee
 we get the conformally improved Hamiltonian
\[
 H = \int dx\ \H = \opergam \int dx\mm\left\{
 	\Pi^2 + \quarter \Phi'^2 + \e{-2 \i \Phi}  - \i \Phi'' \right\}
\]
 and  the improved momentum
\[
 P = \int dx\ \P = \opergam \int dx\mm\left\{
 	\Phi' \Pi - \ipert\Pi' \right\} \.
\]
 According to the prescriptions of radial quantization \cite{bpz} we
know that on a cylinder the role of the lightcone energy-momentum
tensor is played by the sum of energy and momentum densities
\be
 s_+(x) = \H(x) + \P(x) \. \mabel{esplus}
\ee
 This sum generates the current algebra
\[
 \{s_+(x), \mm s_+(y)\} = 2 \mm (s_+(x) + s_+(y))
 \mm\delta'(x-y) \mmplus \opergam \delta'''(x-y)
\.
\]
 Upon Fourier expanding the light-cone energy-momentum, we get a copy
of the classical version of the Virasoro algebra,
\[
 \i \{L_n, L_m \} = (n-m)L_{m+n}
  - \frac{\pi}{2\gamma}(n^3 - n)\delta_{n+m,0} \,
\]
 from which we read off the classical central charge
\[
 c= - 6 \pipergam \.
\]
 The difference $s_-(x)
= \H(x) -\P(x)$ generates another, commuting copy of this algebra.

 In \cite{curtho} this system was canonically quantized using a normal
ordering prescription to cope with divergences. The
quantum corrections shifted the central charge to
\[
c= 1-6(\pipergam + \gamperpi -2) \,
\]
 which yields minimal theories for $\gamperpi$ rational, i.e. when the
deformation parameter $q=\exp{i\gamma}$ is a root of unity. The subset
$\gamperpi = \frac{\nu}{\nu+1}, \ \ \nu=2,3,\ldots$ corresponds to
unitary theories. Similar results were obtained in \cite{gernev}.

\section{An L-matrix for Liouville Theory}

Instead of normal ordering, we shall regularize the ultraviolet
divergencies by putting Liouville theory on  a lattice, in a way that
preserves the integrability of the model.

To get into a position where the methods of the algebraic Bethe
Ansatz (see e.g. \cite{fadBA}) can be used, we have to find a spectral
parameter dependent quantum Lax operator for lattice Liouville theory.
To do this, we shall consider Liouville theory as the massless
limit of sine-Gordon theory. Indeed, the sine-Gordon equation of
motion\footnote{the peculiar choice of normalization of sine-Gordon
mass makes the quantum group structure more transparent in the sequel.}
\[
 \dalembert\Phi + 8 m^2 \sin(2\Phi) = 0
\]
  goes into  the (imaginary) Liouville \eq \refe{lioueq}, if we
rescale the field:
\be
 \Phi \to \Phi + \i\zeta \,
 \mabel{scale}
\ee
 and take the limit
\be
 m\to 0 \; \zeta\to\infty\, \  \ \mbox{ so that}\ \  m\e{\zeta} = 1 \.
 \mabel{limit}
\ee

\subsection{The Sine-Gordon L-matrix}

For lattice sine-Gordon there exists a Lax-operator which is based on
a infinite dimensional representation of the underlying quantum group
\slqt\ \cite{izekor,fadBA}. In this approach, lattice sine-Gordon is
treated as an inhomogeneous XXZ spin chain.

In an auxiliary matrix space $a$ of $2\times2$ matrices, we  define the
$L$-operator  of the $n$:th site of a XXZ-chain to be the  matrix operator
\be
 L_{n,a}^{{\rm xxz}}(\lam) = \mat{ll}{
 \sinh(\lam + \i\gam\mm S^{(n)}_3) &  \i  S^{(n)}_- \mm \singam\\
 \null&\null\\
 \i  S^{(n)}_+ \mm \singam& \sinh(\lam - \i\gam\mm S^{(n)}_3) } \,
 \mabel{Lxxz}
\ee
 a function of the spectral parameter $\lam$.  If the quantum
operators defined on the sites generate the quantum group \slqt,
\be
 \com{S^{(n)}_+}{S^{(m)}_-} = \frac{\sin(2\gam S^{(n)}_3)}{\singam}
\ \delta_{n,m}
\;  \com{S^{(n)}_3}{S^{(m)}_\pm} = \pm S^{(n)}_\pm \ \delta_{n,m} \,
 \mabel{QG}
\ee
 the L-operators \refe{Lxxz} acting on auxiliary spaces $a_1$ and $a_2$
fulfill the fundamental commutation relations (FCR)
\bea
 R_{12}(\lam-\mu) \stack{1}{L}_n(\lam) \stack{2}{L}_n(\mu) &=&
\stack{2}{L}_n(\mu) \stack{1}{L}_n(\lam) R_{12}(\lam-\mu)
 \mabel{FCR} \\
\stack{1}{L}_n(\lam) \stack{2}{L}_m(\mu) &=&
\stack{2}{L}_m(\mu) \stack{1}{L}_n(\lam)\,\ \mbox{for}\ m\neq n\.
 \nonumber
\eea
 Here the usual notation for matrices on the tensor product of two
auxiliary spaces is adopted; $\stack{1}{L}_n \equiv L_{n,a_1}\tensor
\unit_{a_2}$ etc.

The FCR encode the integrability of the system, and they are the basis
of utilizing the algebraic Bethe Ansatz. The XXZ chain belongs to the
class where the R-matrix is trigonometric:
\be
 R=\mat{cccc}{\alpha & \null& \null& \null \\
              \null & \beta& \delta& \null \\
              \null & \delta& \beta& \null \\
	      \null & \null& \null& \alpha}  \;
                               \begin{array}{l}\alpha=\sinh(\lam+\i\gam)\\
					       \beta=\sinh\lam\\
					       \delta=\i\singam \end{array}\.
\mabel{R}
\ee
 The ordered product of L-matrices around the periodic chain is the
monodromy matrix
\be
 T(\lam) = L_{\sN,a}(\lam)\mm L_ {\sss{ N-1},a}(\lam)
\mmdots L_{\sss{1},a}(\lam) \equiv \mat{cc}{\A(\lam)&\B(\lam)\\
					    \C(\lam)&\D(\lam)} \.
 \mabel{monodromedary}
\ee
 The conserved quantities can now be expressed as traces (over the
auxiliary space $a$) of powers of $T$, which all commute due to the
\FCR. The trace of $T$ over the auxiliary space can be interpreted as
the row-to-row transfer matrix of the corresponding two-dimensional
statistical model,
\be
 \tau(\lam) = \mbox{Tr}_a\bigl(T(\lam)\bigr) = \A \mmplus \D \.
\mabel{transfer}
\ee

\vali

To get a $L$-matrix for sine-Gordon, we use an infinite dimensional
representation of the quantum group, generated by the canonical
variables $\Phi_n,\mm \Pi_n$ with commutation relations
\be
 \com{\Phi_n}{\Pi_m} = \i\gam\mm\delta_{mn} \.
 \mabel{commutator}
\ee
 Now we can write  the generators
\be
 S_3^{(n)} = - \opergam\Phi_n \;   S_\pm^{(n)}
  = \frac{1}{2 m \singam} \e{\pm\ipert\Pi_n}
\Bigl(1 + m^2\e{2\i\Phi_n}\Bigr) \e{\pm\ipert\Pi_n} \,
 \mabel{QGgen}
\ee
 which fulfill the commutation relations \refe{QG}.

 Using Generators \refe{QGgen} in \eq\refe{Lxxz} and multiplying with
the matrix $-2 m \i \mm \sigma_1$, we get the $L$-matrix of lattice
sine-Gordon theory
\bea
 L^{SG}_{n,a} &=& -2 m\i\sigmaI L_{n,a}^{{\rm xxz}} \cr
 &\null& \cr &=&  \mat{ll}{
   h_+(\Phi_n)\e{\i\Pi_n} & -2m\i\sinh(\lam + \i\Phi_n) \\
 -2m\i\sinh(\lam - \i\Phi_n)  &   h_-(\Phi_n)\e{-\i\Pi_n}}  \.
 \mabel{sGL}
\eea
 Here we have denoted
\[
 h_\pm(\Phi) \equiv  1 + m^2 \e{\pm 2\i\Phi +\i\gam} \.
\]
 Multiplying $L$ with $\sigma_1$ is a symmetry of the FCR, so
L-matrix \refe{sGL} still satisfies \eq\refe{FCR}.

\subsection{Massless Limit of the  Sine-Gordon L-matrix}

Now we perform the scaling and limiting procedure of Equations
(\ref{scale},\ref{limit}) on the sine-Gordon L-matrix \refe{sGL}, in
order to get a $L$-matrix for lattice Liouville theory.

The functions $h_\pm$ scale to
\[
 h_+ \to 1 \; h_- \to 1 - \e{-2\i\Phi + i\gam} \equiv  h \.
\]
 Performing
(\ref{scale},\ref{limit}) directly on \refe{sGL}, we thus get
\be
 \tilde L_{n,a}^{\L} =  \mat{ll}{
 \e{\i\Pi_n} & -\e{-\lam - \i\Phi_n} \\
 \e{\lam - \i\Phi_n}  &    h(\Phi_n)\e{-\i\Pi_n}}  \. \mabel{trivL}
\ee
 This $L$-matrix with spectral parameter was acquired in
\cite{fadtak}. The $\lam$-dependence of $ \tilde L_{n,a}^{\L}$ is
trivial, though, it can be removed by a lattice gauge
transformation. In other words, the quantum determinant (the central
element of the algebra generated by the elements of $\til L_{n,a}$) is
independent of $\lambda$.  If the $\lam$ dependence is removed, we are
left with the constant $L$-operators inherent in the approaches of
\cite{fadtak,gervais,babelon}. As such, this $L$-matrix is not
viable for the Bethe Ansatz. However, let us comment that work with Lax
operator {\refe{trivL}} leads naturally to the lattice deformation of
Virasoro algebra  \cite{fadtak,volkov,fadvol,babelon}.

To get a Liouville $L$-matrix with non-trivial spectral parameter
dependence, we have to manipulate \refe{sGL} in a more involved
way.\footnote{This prescription was communicated to us by A. Volkov.}
First, we perform a constant lattice gauge transformation on
\refe{sGL}:
\[
 L^{SG}_{n,a } \to g \mm L^{SG}_{n,a} \mm g^{-1} \;
 g = \mat{ll}{m^{\half}&\null\\ \null & m^{-\half} }
\]
 Then we make the scaling \refe{scale}, accompanied by a
renormalization of the spectral parameter:
\be
 \lam \to \lam - \zeta \.
 \mabel{renorm}
\ee
 After scaling and renormalizing, the limiting procedure \refe{limit}
produces
\be
 g \mm L^{SG}_{n,a} \mm\mm g^{-1} \to \mat{ll}{
 \e{\i\Pi_n} & -\i\e{-\lam - \i\Phi_n} \\
 -2\i\sinh(\lam - \i\Phi_n)  &    h(\Phi_n)\e{-\i\Pi_n}}
 \equiv   L^{\L}_{n,a} \.
 \mabel{Lliou}
\ee

This is indeed the sought for quantum $L$-matrix for Liouville theory
with a non-trivial spectral parameter, corresponding to an integrable
lattice regularization of the Liouville system described by Action
\refe{claction}.

The easiest way to see that $L$-matrix \refe{Lliou} corresponds to an
integrable lattice version of quantum Liouville theory, is to take the
classical continuum limit of \refe{Lliou} , and find the corresponding
classical continuum dynamics.

To define the classical limit, one has to recover Planck's constant.
This is achieved by reinterpreting $\gam \to \hbar\gam$ in all quantum
expressions. In the classical limit commutators turn into Poisson
Brackets according to the usual Heissenberg correspondence,
$\frac{\i}{\hbar}[,] \to \{,\}$.

Similarly, to find the continuum limit the lattice spacing $a$ has to
be recovered.  The lattice mass in \refe{sGL} should be $m = a\mm m'
$, and the {\it continuum} mass $m'$ should be taken to zero according
to \eq \refe{limit}.

The continuum variables are defined by
\[
 \Pi_n \to a \mm\Pi(x)\; \Phi_n \to \Phi(x) \;
 \delta_{mn} \to a\mm\delta(x-y) \,
\]
 which maps the discrete brackets corresponding to \refe{commutator}
to the continuous ones of \eq \refe{pbs}.

Now we get in the classical continuum limit the matrix
\be
 U(x,\lam) = \lim_{a, \hbar \to 0} \mm \trac{1}{\i a} \mm
(L^\L - \unit) = -\mat{ll}{
 	-\Pi(x) & \e{-\lam - \i\Phi(x)} \\
	 2 \sinh(\lam + \i\Phi(x))  &   \Pi(x)}  \.
 \mabel{U}
\ee
 As the classical limit of  \FCR\ \refe{FCR}, $U$ satisfies the so called
fundamental Poisson brackets
\[
 \pois{\stack{1}{U}(x,\lam)}{\stack{2}{U}(y,\mu)} =
 \i\gam\com{r_{12}(\lam-\mu)}{\stack{1}{U}(x,\lam) +
 		\stack{2}{U}(x,\mu)} \mm \delta(x-y) \,
\]
  with the trigonometric classical $r$-matrix
\[
 r(\lam) = \lim_{\hbar \to 0} \mm\frac{-1}{\i\hbar} \mm
\left(\frac{R(\lam)}{\sinh\lam} - \unit\right)
 = \frac{-1}{\sinh\lam}\mat{cccc}{
              \cosh\lam & \null& \null& \null \\
              \null & 0& 1& \null \\
              \null & 1& 0& \null \\
	      \null & \null& \null& \cosh\lam} \.
\]

Together with the matrix
\[
 V(x,\lam) = - \mat{ll}{
 	\half\Phi'(x) & \e{-\lam -\i\Phi(x)} \\
	2 \sinh(\lam - \i\Phi(x))  &   -\half\Phi'(x)}   \,
\]
  the classical L-matrix $U$ forms  a Lax-pair for Liouville theory:
\[
 \dot U + V' + i\com{U}{V}  = 0
 \iff \dalembert\Phi - 4\i\e{-2\i\Phi} = 0 \.
\]
 This Lax-pair can be acquired in the scaling and limiting procedure
(\ref{scale},\ref{renorm},\ref{limit}) from the Lax-pair of classical
sine-Gordon in Reference \cite{fadkor}.

A. Volkov brought into our attention the fact that Lax-matrix
{\refe{U}} is gauge equivalent to
\[
 \til U(x,\mu) = \mat{cc}{0& \  \mu - s_+\\
		          1&0} \,
\]
 where $\mu = \exp{2\lam}$, and $s_+$ is the energy-momentum density of
\eq\refe{esplus}. The Lax Equation turns into the Schr\"odinger
equation
\be
 -\psi'' + s_+\psi = \mu \psi \,
 \mabel{scroe}
\ee
 which usually is used in connection to the KdV equation. The role of
\eq\refe{scroe} for Liouville theory is stressed in {\cite{dzpopota}}.

\section{Bethe Ansatz for lattice Liouville theory}

In the \ABA, one tries to triangularize the local $L$-operators in
order to diagonalize the transfer matrix  over the
chain, which yields the conserved quantum quantities. This is done by
finding a local pseudovacuum, which is annihilated by one of the
off-diagonal components in $L_{n,a}$. In addition, to get an
eigenstate of the transfer matrix, the pseudovacuum should be an
eigenstate of the diagonal components.

The $L$-operator \refe{Lliou} developed in the previous section does
not have a local pseudovacuum.  However, as is the case with
it's ancestor $L^{SG}$ \cite{izekor}, the product of two $L^\L$:s from
adjacent sites indeed has a pseudovacuum.

We denote $\Phi_{2n} = \Phi_2\mm ;\ \Phi_{2n-1} = \Phi_1$, and
similarly for $\Pi$.  The product of two Lax-operators is thus
\be
 \L = L^\L_{2n,a}\mm L^\L_{2n-1,a} \mm =  \mat{cc}{A&B\\ C&D}   \,
 \mabel{LL}
\ee
 with
\beastar
 A &=& \e{\i(\Pi_2+\Pi_1)} -2 \e{-\lam - \i\Phi_2} \sinh(\lam - \i\Phi_1) \\
 B &=& -\i\e{-\lam+\i(\Pi_2 -\Phi_1)}
            -\i h(\Phi_1)\e{-\lam - \i(\Phi_2 +\Pi_1)} \\
 C &=&  -2\i\sinh(\lam -\i\Phi_2)\e{\i\Pi_1}
   - 2 \i h(\Phi_2)\e{-\i\Pi_2}\sinh(\lam - \i\Phi_1) \\
 D &=&   h(\Phi_2)  h(\Phi_1) \e{-\i(\Pi_2+\Pi_1)}
  - 2 \sinh(\lam -\i\Phi_2)\e{-\lam-\i\Phi_1}
\eeastar

Inspired by \cite{izekor,fadBA} we  make the Ansatz
\[
 \psi = f(\Phi_1) \mm\delta(\Phi_1-\Phi_2 -\gam)
\]
 for the pseudovacuum at site $n$, with $f$ a functional to be
defined.  Demanding that the off-diagonal operator $C$ annihilates
the vacuum, we get the functional relation
\be
 f(\Phi+\gam) = - h(\Phi) \mm f(\Phi) \.
 \mabel{functrel}
\ee
 This equation sets very stringent conditions on the function $f$. It
is a  sufficient condition for  the pseudovacuum;
when it is fulfilled, the actions of $A$ and $D$ on $\psi$ are
diagonal, with the eigenvalues:
\bea
 A\mm\psi &=& (\!\e{-2\lam +\i\gam} -1)\mm \psi \equiv a(\lam)\mm\psi \cr
 D\mm\psi &=& (\!\e{-2\lam -\i\gam} -1)\mm \psi \equiv d(\lam)\mm\psi  \.
 \mabel{aetd}
\eea

The treatment of Functional Relation \refe{functrel} depends crucially
of the value of $q~=~\e{\i\gam}$.  When $|q| < 1$, Equation
\refe{functrel} is exactly fulfilled by the quantum dilogarithm, see
Ref. \cite{fadkas}. Following \cite{fadcur}, we get an explicit solution
for the case $|q| =1$ as well, which  is of interest to this paper:
\[
 f(\Phi) = \exp{\mm \int_{-\infty}^\infty
\frac{dx}{4x}\mm\mm \frac{\e{(\gam-\pipert-\Phi) x}}
{\sinh\pipert x \sinh\gampert x}
} \mm\.
\]
  The singularity of the integral at $x=0$ is left under the
integration path.  For more discussion on solutions of \refe{functrel}
and other similar functional equations we refer to
\cite{volkov,fadvol,fadkas,fadcur}.

{}From the local pseudovacuums $\psi_n$ we can now build up a total
pseudovacuum for the quantum chain:
\[
 \Psi = \tensor_{n=1}^N \psi_n \,
\]
 where  the amount of paired sites is denoted by $N$.

The two-site L-operators \refe{LL} become triangular when acting on the
local vacuum. Thus the monodromy \refe{monodromedary} acting on the
total pseudovacuum $\Psi$ is triangular as well:
\[
 T(\lam) \mm\Psi = \mat{cc}{a^N(\lam)\mm\Psi & *\\
			    0 & d^N(\lam)\mm\Psi } \.
\]
 The star in the upper right corner denotes a complicated state
created by various combination of $A$:s, $D$:s and one $B$ acting on
$\Psi$.

Correspondingly, the transfer matrix \refe{transfer} has the
pseudovacuum eigenvalue
\be
 \tau(\lam)\mm \Psi = (a^N(\lam)\mmplus  d^N(\lam))\mm\Psi \.
 \mabel{eigentransfer}
\ee

\subsection{The Bethe Ansatz Equations}

In the \ABA\ one makes the assumption that the exited states of the
theory can be obtained from the pseudovacuum by acting on it with the
``pseudoparticle creation operator'' $\B$.  An arbitrary state of the
form
\be
 \Psi_m = \prod_{j=1}^m \B(\lam_j) \mm\Psi
\mabel{psim}
\ee
 is characterized by $m$ values of the spectral parameter $\lam_j$.
This state is an eigenstate of the trace of the monodromy,
i.e. $\A+\D$, if the spectral parameters $\set{\lam_j}$ satsify a
set of $m$ coupled trancendental equations known as the \BAE.
Written in terms of  the components
of $R$-matrix \refe{R} and the eigenvalues $a^N, d^N$ of $\A$
and $\D$, these equations read
\be
 \left[\frac{a(\lam_k)}{d(\lam_k)}\right]^N =
\prod_{\stackrel{j=1}{ j\neq k}}^m
 \frac{\alpha(\lam_k-\lam_j)\beta(\lam_j-\lam_k)}{\alpha(\lam_j-\lam_k)
                                            \beta(\lam_k-\lam_j)}
 \ \ \forall\ k\.
 \mabel{fBAE}
\ee

 Solutions to \refe{fBAE} appear in sets of $m$ spectral parameters,
so called \BA\ roots, which parameterize different states of the
system. A more refined analysis shows that generically all \BA\ roots
are distinct, and that $m\leq N/2$, see e.g. \cite{fadBA}.

For lattice Liouville theory, the eigenvalues $a$ and $d$
are given by \eq\refe{aetd}, and the ratio on the left hand side of
Equation \refe{fBAE} is
\[
 \frac{a(\lam)}{d(\lam)} = \e{\i\gam} \BAEspsmi{\lam}{\i\gampert} \.
\]
 This leads to the \BAE
\be
 \e{\i N\gam} \left[\BAEspsmi{\lam_k}{\i\gampert}\right]^N =
\prod_{j,\mm j\neq k} \BAEspspl{\lam_k-\lam_j}{\i\gam} \ \ \forall\  k\.
 \mabel{BAE}
\ee
 On the other hand, the \BAE\ of a spin-S XXZ chain are
(see e.g. \cite{fadBA})
\be
  \left[\BAEspspl{\lam_k}{\i S\gam}\right]^N =
 \prod_{j,\mm j\neq k} \BAEspspl{\lam_k-\lam_j}{\i\gam} \ \ \forall\  k\.
 \mabel{xxzBAE}
\ee
  Comparing \refe{BAE} to \refe{xxzBAE} we see that in terms of the Bethe
Ansatz, lattice Liouville theory corresponds to a spin $(-\half)$ XXZ
spin chain, with an extra phase factor $\e{\i N\gam}$.

Phase factors of the form $\e{\i \theta}$ have appeared earlier in the
literature on the \BA. In \cite{karowski} it was used to analyze \BAE\
for Potts models, related to a seam insertion of a boundary
field. Similarly, twisted boundary conditions in Reference
\cite{kluwezi} manifest themselves in the form of extra phase factors.
Most interestingly, in the analysis of the eight-vertex model
\cite{baxter,takfad79}, there appears a ``theta-vacuum like'' phase
factor in the Bethe equations, related to different SOS-sectors of the
theory.

Here, however, the important difference to the situations of
\cite{karowski,kluwezi,baxter,takfad79} occurs that the length $N$ of
the chain appears in the phase factor. When $q$ is a root of unity,
this will give us an extra integer parameter to parametrize
``primary'' exited states, in addition to the string length used in
\cite{karowski}. These two integers then give us enough freedom to
parameterize the whole \Kac table.

Thus different chain lengths occurring in the lattice Liouville model
correspond exactly to the different ``theta-vacuum'' sectors in a
critical (R)SOS model.

\section{Mapping  the Lattice Liouville Model
		to the Spin $\half$ XXZ Chain }\mabel{maptoxxz}

\noindent
The thermodynamics \cite{taksuz} and finite size effects
\cite{devewoy,karowski,kluwezi,boizre} of the \BA\ of the fundamental spin
$+\half$ representation of XXZ-chains have been widely discussed in
the literature. Accordingly it would be a desirable goal to map the
spin $(-\half)$ Liouville Bethe equations \refe{BAE} to the spin
$+\half$ XXZ ones \refe{xxzBAE}.  This can be done in the string
picture \cite{taksuz} of solutions to the \BAE.

In order to do the mapping, we thus make the string hypothesis that in the
thermodynamic limit $N\to\infty$, complex \BA\ roots $\set{\lam_j}$
cluster to complexes of the form
\be
 \lam^{(\sK)} = \lam_\sK \mmplus k \i\gam\, \ \ k= -\half (K-1), \mm\ldots\mm,
 \mm\half(K-1) \; \Im\lam_\sK\in\{0,\pipert\}\, \ \ K\in Z_+ \.
 \mabel{kstring}
\ee
 A collection of $K$ \BA\ roots with a common center obeying
\eq\refe{kstring} is known as  a $K$-string. One-strings are the
usual real \BA\ roots. Strings with $\Im\lam_\sK = 0$ are called
positive parity strings, if  $\Im\lam_\sK = \pipert$ one speaks about
negative parity strings.

It is important to notice that when $q$ is a root of unity there
is an upper limit to the length of the string $K$
\cite{taksuz}. With
\[
 \left[\e{\i\gam}\right]^{\nu+1} = \pm 1 \,
\]
 the maximal string length is
\[
 K_{{\rm max}} = \nu.
\]
 This limit on the range of $K$ changes the usual concept of
combinatorical completeness of \BA\ states.  In Ref. \cite{jutkar} it
was argued for a \slqt\ invariant spin chain at $q$ root of unity
that the \BA\ states exhibit completeness in the space of ``good''
representations of the quantum group.

 From now on we shall concentrate only on the root of unity case. More
specifically, we shall assume that the Liouville coupling constant
takes only the values
\[
 \gam = \frac{\pi\nu}{\nu +1} \; \nu\in Z_+ +1 \.
\]
 Using the string picture, we shall be able to map the states of a
Liouville chain with ``anisotropy'' $\gam$ to a spin $\half$ spin
chain with anisotropy
\be
 \tilgam = \frac{\pi}{\nu +1} \; \nu\in Z_+ +1 \.
 \mabel{tilgam}
\ee

We begin with positive parity strings.  The number of $L$-strings is
$n_\sL$, and the total number of \BA\ roots is
\[
 m = \sum_{L=1}^\nu L \mm n_\sL \.
\]
 The rapidities of different $L$-strings are denoted $\lam_{\sL,j}\, \
j=1\mmdots n_\sL$.

Multiplying the \BAE\ \refe{BAE} for all the rapidities comprising a
$K$-string, we get a set of coupled equations of the {\it
real parts} of the strings only.

The terms in the right hand side of (\ref{fBAE}--\ref{xxzBAE}), which
describe the effect on scattering of pseudoparticles on each other,
now become scattering matrices of strings on strings.

The scattering of $K$-strings on 1-strings is described by the
function
\bea
 S_{\sss{ K 1}}(\lam) & \equiv&
\prod_{k=-\Kmiopert}^\Kmiopert  S_{1 1}(\lam)\ =
\prod_{k=-\Kmiopert}^\Kmiopert  \BAEsps{\lam}{\i k\gam} \cr
& &\cr
& &\cr
&=&  \BAEsps{\lam}{\half(K+1)\i\gam}\
\BAEsps{\lam}{\half(K-1)\i\gam} \. \mabel{Sk1}
\eea
 In terms of this function, we can write the scattering matrix
of $K$-strings on $L$-strings as
\bea
 S_{\sK\sL}(\lam) &=& \prod_{k=-\Kmiopert}^\Kmiopert\
\prod_{l=\Lmiopert}^\Lmiopert\
 \BAEsps{\lam}{(k-l+1)\mm\i\gam} \cr
& &\cr
&=&  \prod_{k=\sss{ \jhalf |K-L|}}^{\sss{ \jhalf
(K+L) - 1}} \    S_{k1}(\lam) \
= \prod_{k=\sss{ \jhalf |K-L|}}^{\sss{ {\rm
min} \left[\jhalf (K+L) - 1,\ \nu -\jhalf(K+L)\right]}} \
S_{k1}(\lam)
\.
 \mabel{S}
\eea
 The first expression in terms of $S_{k1}$ bears close resemblance to
(an exponential form of) the decomposition of the tensor product of
two irreducible $SU(2)$ representations with spins $\half(K-1)$ and
$\half(L-1)$. Indeed, the \BA\ can be viewed as a
different way of reducing tensor products of spin 1/2 particles into
irreducible representations. The factor $S_{\sK 1}$ attached to a
spin $\half(K-1)$  representation is obtained by fusing $K$
factors $S_{11}$ corresponding to spin 1/2 representations. The
$S$-matrix then naturally reflects the decomposition of the tensor
products of the representation spaces involved in the scattering.

The second (``reduced'') expression in terms of $S_{k1}$ is peculiar
to the root of unity case. It may be related to the decomposition of
the tensor product of two ``good'' quantum group representations,
c.f. \cite{jutkar}.  The reduced decomposition is related to the
fusion rules of conformal blocks as treated in {\cite{verlinde}},
suggesting that conformal blocks might be represented by \BA\
strings. In the sequel we will see, how this is to be interpreted.

The reduced form shows explicitly the symmetry of
$S_{\sK\sL}$ which will allow us to map the Liouville Bethe equations
on the spin $\half$ XXZ Bethe equations:
\be
 S_\sss{ \nu+1-K,\mm\nu+1-L}(\lam)
 = S_{\sK,\sL}(\lam) \.
 \mabel{Sklsym}
\ee

 Expressing the Liouville Bethe equations \refe{BAE} in terms of
strings we get
\be
 \e{\i K N\gam} \ \left[\BAEspsmi{\lam_{\sK,i}}{\i\kpert\gam}\right]^N
 = \prod_{\sL; \ n_\sL\neq 0} \
   \prod_{\stackrel{j=1}{(\sL,j)\neq(\sK,i)}}^{n_\sL}
   S_{\sK\sL}(\lam_{\sK,i} - \lam_{\sL,j})\all (K,i)\.
 \mabel{sBAE}
\ee

The  amount of coupled equations is reduced to $\sum_{\sL = 1}^\nu
n_\sL$, and all roots of the equations are now taken to be real.

Similarly, applying the string picture to Equation \refe{xxzBAE} for
spin $\half$, we get the spin $\half$ XXZ \ Bethe equations in terms
of strings,
\be
 \left[\BAEspsmi{\lam_{\sK,i}}{\i\kpert\gam}\right]^N
 = \prod_{\sL; \ n_\sL\neq 0} \
   \prod_{\stackrel{j=1}{(\sL,j)\neq(\sK,i)}}^{n_\sL}
   \Bigl(S_{\sK\sL}(\lam_{\sK,i} - \lam_{\sL,j})\Bigr)^{-1} \all (K,i)\.
 \mabel{xxzsBAE}
\ee
 Notice that as compared to {\refe{xxzBAE}}, we have inverted the equation.

Inspired by Equation \refe{Sklsym}, we parameterize the string lengths
{}from the maximal string backwards,
\[
 K = \nu +1 - \tilde K \; L = \nu +1 - \tilde L\; \tilde K,\tilde L =
1, \mmdots , \nu \.
\]
 With this parameterization, it turns out that the Liouville \BAE\
\refe{sBAE} for the strings $\set{\lam_{\sL,j}}$ maps to the spin
$\half$ XXZ equations for a set of strings $\set{\tillam_{\tilde
\sL,j}}$ with the anisotropy $\tilgam$.

Indeed, after some algebra we have for the left hand side
\be
 \e{\i K \gam}\  \BAEspsmi{\lam_{\sK,i}}{\i\kpert\gam} =
 \e{\i \tilde K \tilgam}\
\BAEspsmi{\tillam_{\tilde\sK,i}}{\i\trac{\tilde K}{2}\tilgam} \.
 \mabel{pmap}
\ee
 The spectral parameter is changed in the following way:
\be
 \tillam = \lam - \i (1+(-1)^\sK) \piperf \, \mabel{tillam}
\ee
 i.e. the parity of even-length strings is changed.  In addition, it
turns out that
\[
 S^\gam_\sss{ K 1}(\lam)
= \Bigl(S^\tilgam_\sss{ \til K 1}(\til\lam)\Bigl)^{-1} \,
\]
 where the upper index denotes whether $\gam$ or $\tilgam$ is
used when defining $S_\sss{K1}$ according to Equation \refe{Sk1}.

 Using this, as well as Symmetry Property \refe{Sklsym}, we get for
the string-on-string scattering matrix \refe{S}
\be
 S_{\sK\sL}^\gam (\lam_{\sK,i} - \lam_{\sL,j})
 =\Bigl(S_{\tilde\sK\tilde\sL}^\tilgam
   (\tillam_{\tilde\sK,i}-\tillam_{\tilde\sL,j})\Bigr)^{-1} \.
 \mabel{smap}
\ee

These results can easily be generalized to  negative parity Liouville
strings as well.

Now we are in position to state the equivalence of lattice Liouville
and \halfxxz\ Bethe Ans\"atze.  Using Equations (\ref{pmap},
\ref{smap}) in \refe{sBAE} and comparing to Equation \refe{xxzsBAE},
we get {\it complete equivalence} of the string states in lattice
Liouville theory and a spin $\half$ XXZ chain with an additional phase
factor $ \exp\{\i N \tilde K \tilgam\}$. This extra factor will allow
us to incorporate the remnant of the chain length mod $\nu+1$ as a
meaningful extra parameter in the theory of \FSC\ of an usual XXZ
chain.

The spin $\half$ XXZ \BAE\ we shall be analyzing are thus
\be
 \e{\i N \til K\tilgam}\mm \left[
\BAEspsmi{\tillam_{\tilde\sK,i}}{\i\trac{\tilde K}{2}\gam}
\right]^N \times
\prod_{\sss{\til L}; \ n_\sss{\til L}\neq 0} \
   \prod_{\stackrel{j=1}{\sss{(\til L,j)\neq(\til K,
  		i)}}}^{n_\sss{\til L}}
   S_{\sss{\til K\til L}}(\lam_{\sss{\til K},i}-\lam_{\sss{\til L},j})
\ = \ 1 \all (\til K,i) \.
 \mabel{phxxzBAE}
\ee
 This equation is valid separately for each $N$, but
the thermodynamic limit $N\to\infty$ is sensible only if $N$
approaches $\infty$ in steps of $\nu+1$.

Accordingly, we parametrize the (even) chain length  as
\be
 \frac{N}{2} = n (\nu +1) + \kap\, \ \ \kap=0,\ldots ,\nu  \.
 \mabel{N}
\ee
 Taking the thermodynamic limit $N\to\infty$ {\it at fixed} $\kap$,
letting $n\to\infty$, the limit of \BAE\ (\ref{sBAE}, \ref{phxxzBAE})
is well defined.  The extra phase factor in the \halfxxz\ Bethe
equations \refe{phxxzBAE} is thus
\be
 \e{2 \i\kap \til K \tilgam} \.
 \mabel{phase}
\ee

\vali

To summarize, the obtained equivalence of string states in lattice
Liouville and spin $\half$ XXZ is the following:

\vali\vali

 \noindent\begin{tabular}{l|c|c}
 \null     & Lattice Liouville &
	\begin{minipage}{3cm}\begin{center}\vspace*{2mm}
Spin $\half$ XXZ
	   \vspace*{2mm}\end{center}\end{minipage} \\ \cline{1-3}
anisotropy &
	\begin{minipage}{3cm}\begin{center}\vspace*{2mm}
$\gam=\frac{\pi\nu}{\nu+1}$
	   \vspace*{2mm}\end{center}\end{minipage} &
$\tilgam = \frac{\pi}{\nu +1}$\\
string lengths & $K$                         & $\tilde K =\nu +1 -K$ \\
spectral
parameters &
	\begin{minipage}{3cm}\begin{center}\vspace*{2mm}
$\lam$
	   \vspace*{2mm}\end{center}\end{minipage}  &
$\tillam=\Re\lam+\i\left(\Im\lam-(1+(-1)^\sK)\piperf\right)$  \\
 \end{tabular}

 \vali\vali
\noindent
For the analysis of the physical vacuum, it is important to note
that strings of maximal length are mapped to 1-strings, i.e. ordinary
\BA\ roots, and vice versa.

We are intersted in the energy spectrum of the Liouville model.  The
auxiliary Lax-operator $L_{n,a}$ which was used for the \BA,
intertwines the $n$:th quantum space and the auxiliary space $sl_2$.
The quantum spaces for lattice Liouville model are copies of $L^2({\rm
R})$. Thus $L_{n,a}(\lam)$ does not degenerate to a permutation
operator at any value of the spectral parameter $\lam$, and it is not
good for investigating lattice dynamics.  To get an integrable
Hamiltonian that generates lattice Liouville dynamics, one would have
to introduce fundamental Lax-operators $L_{n,f}$ that intertwine two
quantum spaces \cite{tatafa}.

Fortunately, we are here only interested in energy and momentum
eigenvalues, so we do not have to investigate the fundamental
$L$-operator. Following {\cite{tatafa}} one can read off the energy
and momentum eigenvalues from the eigenvalues of the diagonal elements
$A$ and $D$ of the {\it auxiliary} Lax-operators $L_{n,a}$.  The
momentum eigenvalues are
\[
 P^\L(\set{\lam_j}) =  \trac{1}{2\i} \sum_{j=1}^m  p(\lam_j) \;
 p(\lam) =    \ln \frac{a^\gam(\lam)}{d^\gam(\lam)} \.
 \]
  The upper index for $a$ and $d$ again stresses the particular
value of anisotropy used.

The energy can be acquired by differentiating:
\[
  E^\L(\set{\lam_j}) = \sum_{j=1}^m \epsilon\mm(\lam_j)\;
  \epsilon\mm(\lam)  = \frac{\gam}{\pi}\mm\frac{d}{d\lam}\mm p(\lam)      \.
\]

{}From {\refe{phxxzBAE}} we see that for the  \halfxxz\ with extra phase
factor, the roles of $a$ and $d$ are interchanged. Accordingly,
using  Correspondence {\refe{pmap}}, we can write the Liouville energy
and momentum in terms of the  XXZ ones:
\bea
 P^{\L}(\set{\lam_j})
 &=& \trac{1}{\i}\sum_{j=1}^m
		\ln\frac{a^\tilgam(\lam_j)}{d^\tilgam(\lam_j)}
  = - P^{{\rm xxz}}(\set{\lam_j})
 \equiv
 \i\ln\Lam^{{{\rm xxz}}}(\set{\lam_j})\longat{\lam=\i\trac{\tilgam}{2}}
  \mabel{moment-mal} \\
  E^\L(\set{\lam_j}) &=& -  E^{{\rm xxz}}(\set{\lam_j})
   \equiv -\i\frac{\tilgam}{\pi}\mm \frac{d}{d\lam}\mm
  \ln\Lam^{{{\rm xxz}}}(\lam,\set{\lam_j})
             \longat{\lam=\i\trac{\tilgam}{2}} \ \ \.
 \mabel{energy-mal}
\eea
 Here we have expressed the energy and momentum in terms of
eigenvalues $\Lam$ of the transfer matrix \refe{transfer},
\be
\left(\A(\lam)+\D(\lam)\right)\mm\Psi_m(\set{\lam_j}) \equiv
\Lambda(\lam,\set{\lam_j})\Psi_m(\set{\lam_j}) \.
 \mabel{Lam}
\ee
 This is possible for the \halfxxz\, as the auxiliary
and fundamental Lax-operators for the \halfxxz\ coincide.  The XXZ
Lax-operator \refe{Lxxz} yields local commuting quantities at the
value $\lam=\i\frac{\til\gam}{2}$, for which it becomes a permutation
matrix.

The Bethe equations do not have to be completely solved to get the low
lying spectrum of the Hamiltonian.  We need only the {\it finite size}
i.e. $\operN$ corrections to the eigenvalues of the spin $\half$ XXZ
transfer matrix corresponding to \BAE\ {\refe{phxxzBAE}}. The reason
for this is the following:

In this paper we started by discretizing Liouville theory in a finite
volume $2\pi = a\mm N$, with lattice spacing $a$. Thus the
conformal scaling limit of the spin chain corresponds to the continuum
limit of the Liouville field theory.  Moreover, the continuum
Hamiltonian is
\be
H_{{\rm cont}} = \trac{1}{a}\mm H_{\mm{\rm lattice}}   \,
\mabel{Hcont}
\ee
 so that only $\operN$ corrections to the eigenvalues of the lattice
Hamiltonian remain finite in the continuum limit.  This is exactly the
realm of finite size effects, which in this case are
simultaneously finite lattice spacing effects.

\section{Finite Size Corrections for the six-vertex model}

As is well known, the \halfxxz\ has intimate connections to the
six-vertex model of classical statistical mechanics {\cite{baxter}}.
For two dimensional classical statistical models, the $\operN$
behaviour and the conformal properties are closely related.

In \cite{blocani,affleck,cardy} it was argued that the central charge
$c$ of the conformal field theory corresponding to the scaling limit
of a two-dimensional statistical model is related to the finite size
corrections to the free energy, i.e. the logarithm of the maximal
eigenvalue $\Lam_o$ of the transfer matrix, in the limit $N\to\infty$.
On the other hand, the statistical mechanics minimum free energy
configuration corresponds to the ground state energy $E_o$ of the
scaling conformal theory.  For a model on an infinitely wide strip of
length $N$, the behavior of $E_o$ is \cite{cardy}
\be
 E_o = N f_\infty \mm
   - \mm \operN \trac{\pi}{6}\mm c \mmplus \O(\trac{1}{N^2}) \,
 \mabel{scalec}
\ee
   with $f_\infty$ the free energy per site in the thermodynamic
limit.

Similarly, the critical indices (conformal weights) $\Delta,\Bardelta$
of operators corresponding to exited states of the system can be read
off the large $N$ behavior of configurations close to the one
minimizing the free energy.  In terms of the higher energy and
momentum eigenvalues of the corresponding 1+1 dimensional quantum
theory, critical indices are:
 \bea
E_m - E_o &=& \trac{2\pi}{N}\mm (\Delta \mmplus \Bardelta)
 		\ +\ \O(\trac{1}{N^2}) \cr
 & &   \mabel{scaleweights}\\
P_m - P_o &=& \trac{2\pi}{N}\mm (\Delta \mm - \mm \Bardelta)
 		\ +\ \O(\trac{1}{N^2}) \nonumber
\eea
 As opposed to our differential dependence {\refe{energy-mal}},  the
approach  of Ref. {\cite{cardy}} relates  the energy and momentum
directly to lower eigenvalues $\Lam_m$ of the transfer matrix
at criticality; $E_m\sim -\Re \ln\Lam_m$, $\ P_m\sim -\Im \ln\Lam_m$.

\vali

The \FSC\ of six-vertex models and the corresponding conformal
properties have been extensively studied in the literature
\cite{devewoy,karowski,kluwezi,boizre}.  For the six-vertex model with an
extra phase factor of the form \refe{phase}, and string exitations,
the finite size corrections were calculated by Karowski
\cite{karowski}.  The results of \cite{karowski}  relevant for us are
the following.

The ground state of a six-vertex model is described by a filled Dirac
sea of one-strings, i.e. $n_1= N/2$, $n_l = 0, \mm l>1$.  The
logarithm of the corresponding maximal eigenvalue of the transfer
matrix is \cite[Eq. 4.3]{karowski}\footnote{Note that our $\nu$
differs from the one of \cite{karowski} by one. The relation between
our $\lam$ and the $\theta$ of \cite{karowski} is $\lam = -\i\theta +
\i\trac{\tilgam}{2}$. This follows from the differences of the
respective Bethe equations \refe{phxxzBAE} and \cite[Eq.
2.4]{karowski}.  }
\be
  \ln \Lam_o \approx - \i N f_\infty(\lam) + \frac{1}{N} \frac{\pi}{6}
\left(1 - \frac{6\kap^2}{\nu(\nu+1)} \right) \mm
 \cosh\left(\trac{\pi}{\tilgam}\lam\right) \.
 \mabel{cscale}
\ee

Exited states consist of higher strings above the vacuum, and holes in
the distribution of one-strings.  Low-energy excitations have both the
number of holes and the number of higher strings $\sum_{\sL > 1}
n_\sL$ of the order $N^0$.

For a given distribution of strings $\{n_l\}$, there is a certain
number of allowed values for the spectral parameters.  This number
depends on the behavior of the \BAE\ in the limits $\lam\to\pm\infty$.
Due to the dependence of these limits on higher strings, each
$L$-string gives automatically rise to $2L-2$ holes.

The ``primary'' exitations correspond to states with one higher
string, no extra holes and the holes corresponding to the string
evenly distibuted between the two surfaces of the Dirac sea of
1-strings, i.e. between $\lam \sim\infty$ and $\lam \sim-\infty$.

For such states, the finite size behavior of the eigenvalues of the
transfer matrix reads\cite[Eq.  4.5]{karowski}
\be
  \ln\Lam_m - \ln\Lam_o \approx  -\trac{2\pi}{N}
\left\{(\Delta + \Bardelta)
 \cosh\left(\trac{\pi}{\tilgam}\lam\right)
\mmplus (\Delta - \Bardelta)
 \sinh\left(\trac{\pi}{\tilgam}\lam\right) \right\}\,
 \mabel{weightscale}
\ee
  where the weights are, if the single higher string is a $\til
K$-string,
\be
 \Delta = \Bardelta = \frac{\Bigl((\nu+1)(\til K-1) +\kap\Bigr)^2 -
\kap^2}{4\nu(\nu+1)} \.
  \mabel{delta}
\ee
 If more strings and /or holes are present in the exited state, the
resulting critical indices differ from the ones above by additional
integers.  Accordingly, these states belong to the conformal towers of
descendants of the described ``primary states''.

 For future use we define the function
 \[
  \delta_{L,\kap} = \frac{\Bigl((\nu+1)(L-1) +\kap\Bigr)^2}
    {4\nu(\nu+1)} \,
\]
 in terms of which the critical indices \refe{delta} read
 \be
\Delta = \Bardelta = \delta_{\til K,\kap} - \delta_{1,\kap}\.
 \mabel{Deleqdeldel}
\ee
 In this form, we explicitely see the subtraction of the part
corresponding to the ground state.

An extra condition on $\til K$ for a primary state was found in
\cite{karowski}. There is an upper limit for the length of the strings
that contribute to the \FSC. For extra phase $\kap$, only strings
satisfying
\be
 \til K < \nu + 1 - \kap
\mabel{Klim}
\ee
 contribute. This cutting off of higher strings resembles a result of
Ref. \cite{jutkar} for a \slqt\ invariant XXZ chain with fixed
boundary conditions. There it was conjectured that highest strings
correspond to ``bad'' representations of the quantum group (i.e.
representations with vanishing $q$-dimension), which have to be
removed from the spectrum to keep the theory unitary.

\vali

In the six-vertex model, one expects conformal invariance at $\lam =
0$, where the $R$-matrix becomes isotropic. At this point, one can use
Equations (\ref{scalec}, \ref{scaleweights}) to recognize the conformal
properties of the model.

{}From \eq\refe{cscale} we read off the central charge:
\be
 c =  1 - \frac{6\kap^2}{\nu(\nu+1)} \.
 \mabel{c}
\ee
 For the value $\kap=1$, related to critical Potts models in
\cite{karowski}, the central charges of unitary minimal models emerge.
The critical indices \refe{delta} reproduce a row in the \Kac\ table.
The highest string does not contribute to the spectrum, due to
Restriction \refe{Klim}.

\section{Conformal Properties of Lattice Liouville Theory}

Now we can use the results of Ref. \cite{karowski} reviewed in the
previous section, to calculate the scaling properties of lattice
Liouville energies and momenta, and to recognize the corresponding
conformal structures.

We are interested in \BA\ states that correspond to the ``primary''
states of the six-vertex model, i.e. one $K$-string above the physical
vacuum, with $K = \nu + 1 - \til K$.

{}From the scaling forms of the transfer matrix
(\ref{weightscale}, \ref{scalec}) we get Liouville momenta and energies
using Prescriptions (\ref{moment-mal}, \ref{energy-mal}).  The
resulting eigenvalues yield critical indices using
\eq\refe{scaleweights}.

For lattice Liouville theory, there is an important difference to the
case treated in \cite{karowski}. The extra phase $\kap$ is not a
constant. Instead we have sectors with different values of $\kap$,
corresponding to different lengths of the chain mod $(\nu+1)$, as
indicated by \eq\refe{N}. In the thermodynamic limit, all values of
$\kap$ coexist.

{}From \eq\refe{cscale} it is easy to see, that the ground state of the
theory lies in the $\kap = 0$ sector, which gives $c=1$, the usual
central charge for a periodic XXZ spin chain \cite{devewoy,alibaba}.

As in \cite{karowski}, minimal models would emerge if the ground state
was taken to be in the $\kap=1$ sector. Here we will adopt this
approach, discarding the $\kap=0$ sector alltogether. Later on we will
return to the interpretation of this reduction of the theory.

Accordingly, the central charge is
 \be
c = 1 - \frac{6}{\nu(\nu+1)} \; \nu = 2,3, \ldots
 \mabel{cIII}
\ee
 and we recover the unitary minimal \CFTs\ of \cite{bpz,fqs}.

Calculating the exitation energies and the corresponding critical
indices from \eq\refe{weightscale}, we have to subtract the ground
state energy, not only the minimal energy $\delta_{1,\kap}$ within
each $\kap$ sector.

With the ground state lying in the $\kap=1$ sector,  we get for the
critical indices, instead of (\ref{Deleqdeldel}),
\be
  \Delta=\Bardelta = \delta_{\til K,\kap} - \delta_{1,1} =
\frac{\Bigl((\til K - 1) (\nu+1) + \kap\Bigr)^2
		- 1}{4\nu(\nu+1)}
\. \mabel{weightsIII}
\ee
 Defining
\bea
 p &=& \til K + \kap -1\; p = 1, \ldots, \nu -1 \cr
 q &=& \kap\; q = 1, \ldots, \nu
 \mabel{pq}
\eea
 we can write the critical indices in the form
\be
 \Delta = \Bardelta \ =\ \frac{\Bigl( p\mm (\nu+1) - q\mm \nu\Bigr)^2 -
1}{4\nu(\nu+1)} \.
 \mabel{weightsIV}
\ee
 These scaling weights reproduce the whole \Kac table of unitary
\CFTs. The ranges of $p$ and $q$ follow accurately from Restriction
\refe{Klim} on the maximal string length, and the possible values of
$\kap$ with the $\kap=0$ sector excluded.

\vali

{}From \refe{Hcont} we see that in the continuum, the energy and
momentum of primary
\BA\ states are
\[
 E = \Delta \mmplus \Bardelta\; P = \Delta \mm - \mm \Bardelta \.
\]
 The primary Bethe states are thus products of holomorphic and
antiholomorphic vectors (right and left movers) with conformal weights
\refe{weightsIII}.

\vali

The behavior encountered here is exactly the same as the one
encountered in the restriction of SOS models to RSOS models
\cite{anbafo}.  The \BAE\ \refe{phxxzBAE} are the Bethe equations of
the so called III/IV critical limit of the SOS and corresponding eight
vertex and XYZ models, in the case of root of unity anisotropy.  In
the III/IV critical limit, the extreme off-diagonal term (usually
denoted $d$) in the eight-vertex $R$-matrix vanishes, and the elliptic
functions degenerate to trigonometric ones.  Thus (on applying the
string picture,) the XYZ Bethe equations of \cite{baxter,takfad79}
turn into Equations \refe{phxxzBAE} at criticality.  For an anisotropy
of the form \refe{tilgam}, the SOS Bethe states are parameterized by
all $0\leq \kap \leq \nu$. The ground state lies in the $\kap=0$
sector, and the corresponding central charge is $c=1$.

In the root of unity case, it becomes possible to ``restrict'' the SOS
model \cite{anbafo}. The sectors with $1\leq\kap\leq\nu$ decouple from
the other sectors, and we can restrict our interest to these sectors
only.  This decoupled part of the SOS model is known as the RSOS
model.  On the level of Bethe equations, the restriction means leaving
out the $\kap=0$ sector. The ground state now lies in the $\kap=1$
sector, and in the thermodynamic limit the unitary \CFTs\ with $c<1$
emerge \cite{huse}.

\vali

There is one more subtlety in the interpretation of the results for
the lattice Liouville model described above. Remembering the
correspondence between lattice Liouville theory and the \halfxxz\
described in Section \ref{maptoxxz}, it is evident that the physical
vacuum for Liouville theory consists of maximal strings. This can be
viewed as the maximal string limit of the fact that the vacuum of
higher spin chains consists of higher strings.

Due to the complicated structure of the vacuum, not all
combinations of remnant and string length are a priori allowed.
On the contrary, we get stringent conditions on $N$ from requiring the
coexistence of a specific remnant $N \mod (\nu+1)$ and a single
$K$-string (corresponding to a spin $\half\ $ XXZ $\til K$-string)
above a sea of maximal $\nu$-strings.  In fact, these requirements fix
the chain length modulo $\nu(\nu+1)$.  The parameterization \refe{N}
of $N$ has to be extended to
\be
 \frac{N}{2} =  (n(\nu+1) \mm-\mm \kap \mmplus K)\nu + K
	     =  (n\mm\nu \mm-\mm \kap \mm + \mm  K)
			(\nu +1) + \kap \.
 \mabel{nKr}
\ee
{}From here we see that for this $N$ it is indeed possible to define a
state with one $K$-string over a sea of $n(\nu+1) - \kap + K$ maximal
strings. In addition the remnant is $\kap$. The thermodynamic limit
has to be taken in steps of $\nu(\nu+1)$ by taking $n\to\infty$ in \eq
\refe{nKr}.

Accordingly, the full picture of lattice Liouville primary states is
the following.  In a lattice Liouville chain consisting of $N$ sites
($N$ even), there is a primary state characterized by two integers.
These integers are related to the remnants of the chain lenth $\mod
\nu(\nu+1)$ and $\mod (\nu+1)$,
\[
 \frac{N}2 \mod \nu(\nu+1) = (\nu - p)(\nu +1) + q \.
\]
 The primary state is the state with a single lower string.  The $q=0$
sector exhibits the behavior of the SOS ground state, and after the
restriction, the RSOS unitary series emerge, with central charges
\refe{cIII} and conformal weights \refe{weightsIV}.  In the
thermodynamic limit all remnants mod $\nu(\nu+1)$ coexist (except
possibly for the decoupled $q=0$), and the Liouville primary states
give the whole \Kac table.

\section{Conclusions}

We have developed a spectral parameter dependent integrable structure
to quantum Liouville theory on a lattice. Using the ensuing
$L$-matrix, we have written the \BAE\ for Liouville theory.

We have concentrated on certain Liouville coupling constants $\gam$,
for which $q=\e{\i\gam}$ is a root of unity.  Using the string picture
to describe exited \BA\ states, we have mapped the Liouville Bethe
equations to a set of generalized spin $\half$ XXZ \BAE, more exactly
the critical SOS Bethe equations.  This mapping takes maximal
Liouville strings to XXZ one-strings and vice versa. The physical
Liouville vacuum thus consists of a Dirac sea of maximal strings.

Using results of Karowski \cite{karowski} for the \FSC\ to the
eigenvalues of the transfer matrix, we have calculated the central
charges and conformal dimensions of the \halfxxz, and accordingly also
of Liouville theory.

We found that the continuum limits of lattice Liouville theories with
coupling constants $\gam = \pi\trac{\nu}{\nu+1}\mm, \ \nu=2,3,\ldots$
reproduce the unitary minimal models of Friedan, Qiu and Shenker
\cite{fqs}, on restricting the chain length not to be divisible by
$\nu+1$. This restriction is the exact analogue of the RSOS
restriction of SOS models at root of unity anisotropies.

Primary excitations of the Liouville chain are characterized by two
integers, the length of a shorter string above the vacuum consisting
of maximal strings, and the remnant $\kap$ of the chain length mod
$(\nu+1)$. With these two parameters, the conformal weights
corresponding to the exited states give all states in the
corresponding \Kac table.

To clarify the structure of the different sectors in the theory, a
unitarity analysis based on the hidden \slqt\ symmetry is needed. This
should illuminate both the RSOS restriction $\kap\neq 0$, and the
result of \cite{karowski} that highest XXZ (i.e. lowest Liouville)
strings do not contribute to the spectrum. Following \cite{jutkar} we
believe that these properties are deeply related to properties of root
of unity representations of \slqt. Truncation of the \BA\ Hilbert
space corresponds to excluding ``bad'' representations of the quantum
group, which is  required in order to have a positive metric on
the Hilbert space.

In this work we found equivalence of the \BAE\ of the lattice
Liouville  and  the critical eight-vertex (SOS) models.
Accordingly, the results presented here can be used to provide a
Lagrangian description of the critical conformal field theories of all
two-dimensional statistical models related to the eight-vertex model.

Due to the intimate connection of Liouville theory to two dimensional
gravity, it would be very interesting to extend the method of
quantizing Liouville theory presented here to more general couplings.
It is evident that negative couplings $\gam$ correspond to the real
sector of the Liouville model with $c>25$.

Whether the strong coupling results of Gervais \& al. \cite{gervais}
in the regime $1<c<25$ can be reproduced and \ or extended by \BA\
methods, remains to be seen. For this, one should analyze Liouville
theories with imaginary couplings $\gam$. This requires formulating
the problem in terms of Baxter's equation \cite{baxter}, and use of
quantum separation of variables in Sklyanin style \cite{sklyanin}.

\vali

\subsubsection*{Acknowledgments}

We thank A.G. Izergin, R. Kashaev, V. Tarasov and A. Volkov for
illuminating discussions. In addition, O.T. is indepted to I. Affleck
for a discussion.

The work of LDF was supported by a grant of the Finnish academy and
grant R2H000 of ISF. The work of OT  was supported by a NSERC grant.


\setlength{\baselineskip}{14pt}

\end{document}